# Contact resistivity due to oxide layers between two REBCO tapes

Jun Lu[1,*], Yan Xin[1], Eric Lochner[2], Kyle Radcliff[1], and Jeremy Levitan[1]

[1]*National High Magnetic Field Laboratory, Tallahassee, FL 32310*

[2]*Department of physics, Florida State University, Tallahassee, FL 32306*

Corresponding author: junlu@magnet.fsu.edu

## ABSTRACT

In a no-insulation (NI) REBCO magnet, the turn-to-turn contact resistivity ($\rho_c$) determines its quench self-protection capability, charging delay time and the energy loss during field ramps. Therefore it is critically important to be able to control a range of $\rho_c$ values suitable for various NI magnet coils. We used a commercial oxidizing agent Ebonol® C to treat the copper surface of REBCO tapes. The copper oxide layer was characterized by cross-sectional transmission electron microscopy (TEM) and x-ray photoelectron spectroscopy (XPS). The oxide layer formed in Ebonol® C at 98 °C for 1 min is $Cu_2O$ of about 0.5 μm. The $\rho_c$ between two oxidized REBCO is in the order of 35 mΩ-cm$^2$ at 4.2 K which decreases slowly with contact pressure cycles. The $\rho_c$ increases but only slightly at 77 K. We also investigated the effect of oxidation of stainless steel co-wind tape on $\rho_c$. The native oxides on 316 stainless steel tape as well as those heated in air at 200 - 600 °C were examined by TEM and XPS. The native oxides layer is about 3 nm thick. After heating at 300 °C for 8 min and 600 °C for 1 min, its thickness increases to about 10 and 30 nm respectively. For the stainless steel tapes with about 10 nm surface oxides, pressure cycling for 30,000 cycles decreases $\rho_c$ by almost 4 orders of magnitude. Whereas at 77 K, it only changes slightly. For a surface with 30 nm oxide, the $\rho_c$ decreases moderately with load cycles. The results suggest that for an oxidized stainless steel to achieve stable $\rho_c$ over large number of load cycles a relatively thick oxide film is needed.

1. Introduction

The no-insulation (NI) REBCO magnet has advantages of self-quench-protection and high engineering current density over its insulated counterpart [1]-[4]. But it also has a few significant drawbacks, such as long charging/discharging delays and high ramp losses, which limit its applications in user magnet systems where frequent field ramps are required. In addition, high transient electrical currents during a magnet quench result in high electromagnetic stresses that could damage the magnet [5]. All these drawbacks can be largely mitigated by increasing coils' turn-to-turn contact resistivity $\rho_c$. On the other hand, a very high $\rho_c$ compromises the coil's self-quench-protection ability and results in magnet burn-out during a quench. Since the optimal $\rho_c$ is coil specific, it is highly desirable to develop a technology to control $\rho_c$ in a wide range.

Contact resistivity $\rho_c$ have been measured in short samples and NI REBCO coils by a number of research groups [6]-[9]. Previously we measured $\rho_c$ of short samples at 77 K and 4.2 K under different contact pressures [6]. The control of $\rho_c$ by coating the surface of REBCO tapes with various materials has been attempted; and the effect of cyclic contact pressure was studied [7]. But a technique that reliably controls a wide range of $\rho_c$ values which is consistent over a large number of pressure cycles is yet to be developed.

The theory of electrical contact resistance has been well established [10, 11]. For a contact with a thin resistive film, $\rho_c$ can be written as,

$$\rho_c = \frac{\rho}{2aN} + \frac{\rho_f d}{\pi a^2 N} \tag{1}$$

where $\rho$ and $\rho_f$ are bulk resistivity of the base material and the film respectively; $a$ and $N$ are respectively the average radius and number density of contact asperity spots; $d$ is the thickness of the resistive film. The first term represents the constriction resistance of the base material. The second term is the contribution from the resistive film. In principle, $\rho_c$ can be controlled by changing the thickness $d$ and resistivity $\rho_f$ of the film. In practice though, $a$ and $N$ are unknown, and $\rho_f$ at low temperatures is also difficult to control. So equation (1) can only provide a general guidance for $\rho_c$ control. A reliable $\rho_c$ control method will have to be developed largely by trial and error.

Copper surface exposed to ambient environment has a layer of native oxides which is only a few nanometers thick containing $Cu_2O$ and $CuO$ [12, 13], both semiconductors and highly resistivity at low temperatures [11, 14, 15]. But due to the thin thickness of the native oxides layer, the $\rho_c$ of as received REBCO tapes whose contact is via copper stabilizer layer is low, in the order of $10^{-5}$ $\Omega$-cm$^2$ at 77 and 4.2 K [6]. The existence of the native oxide is consistent with the considerable decrease in $\rho_c$ of REBCO tapes at 77 K after 20,000 pressure cycles, which was interpreted as wearing/penetration of the native oxide layers [7].

In order to obtain a higher $\rho_c$, a thicker oxide layer is needed. Copper oxide growth by heating copper in air does not seem to be a good option. Because the oxide growth rate is very low below 200 °C, while heating REBCO tapes at above 200 °C causes critical current degradation [16]. A better option would be surface treatment by an oxidizing chemical agent at a temperature well below 200 °C. It is known that copper surface can be oxidized by a solution of sodium hydroxide (NaOH) and sodium nitrite ($NaClO_2$) at elevated temperatures [17, 18]. This process has been commercialized, and a copper blackening chemical called Ebonol® C is commercially available at Enthone Inc. The exact reaction chemistry between copper and Ebonol® C is not very clear. It was used to treat NbTi Rutherford cables in order to increase inter-strand contact resistance and reduce ac losses [19, 20]. It was also proposed to be used to treat REBCO tapes to increase $\rho_c$ for NI magnet applications [21, 22].

Another method to control $\rho_c$ in a NI magnet is to control the surface oxides on co-wind tapes. Since high field REBCO magnets often require co-wind tapes for mechanical reinforcement [4, 23], it is convenient to control the surface of co-wind tapes instead of that of the REBCO tapes. This way, the potential damage to the REBCO tapes during handling and the oxidation process is avoid. Compared with the damages to co-wind tapes, the damages to REBCO tapes are much more difficult to identify and much more expensive to replace.

Austenitic stainless steels also have a thin layer of native oxides in ambient environment, which is 2 - 3 nm thick [24]. The native oxides layer consists of $Cr_2O_3$, $Fe_2O_3$ and $Fe_3O_4$ [25]. Upon heating in air, the oxide layer becomes thicker and its composition changes slightly as well [26]. The stainless steel tapes received from different manufacturers can have very different surface conditions depending on the details of the manufacturing process. This may partially explain the considerable difference in $\rho_c$ reported by different groups [7, 27, 28]. Evidently surface oxides and their effects on $\rho_c$ is critically important, but they have not been characterized in the context of $\rho_c$ control.

In this paper, we characterized the oxide layers on REBCO tapes treated by Ebonol® C and on 316 stainless steel by heating in air by using transmission electron microscopy (TEM) and x-ray photoelectron spectroscopy (XPS). The $\rho_c$ of these oxidized samples were measured at 77 K and 4.2 K under contact pressure up to 25 MPa and pressure cycles up to 30,000.

## 2. Experimental

The REBCO tape is SuperPower SCS4050 which has 20 μm plated copper stabilizer. 20% Ebonol® C special (Enthone Inc.) solutions were used to oxidize the copper surface of REBCO tapes. The oxidation was at 98 °C. For copper oxide growth rate studies, we used C110 copper stamp samples of about 12.7 x 15 x 0.5 $mm^3$. Their cold-rolled surfaces were slightly polished by a Scotch-Brite pad. Each sample was weighed before and after oxidation, then weigh again after the oxide was removal by a 3.7% HCl solution for 2 min. The 3.7% HCl does not etch copper as verified by our experiments. A Mettler-Toledo Analytical balance XS104 (max. 120 g, precision 0.1 mg) was used to weigh the samples. The weight losses were used to calculate the oxides thickness. In the calculation, 6.0 and 8.96 $g/cm^3$ were used for densities of $Cu_2O$ and Cu respectively.

The stainless steel tape samples are type 316. They are 4 mm wide and 50 μm thick and in full-hard condition. Since the initial surface condition of the stainless steel tapes is unknown, all the as-received samples were etched by 37% HCl for 5 min to remove the initial surface oxides and reset the surface condition. Subsequently the stainless steel tape samples were heat treated in air at temperatures between 200 and 600 °C in a horizontal quartz tube furnace.

The oxidized surfaces of both REBCO and stainless steel were characterized by transmission electron microscopy (TEM) and x-ray photoelectron spectroscopy (XPS). The TEM was performed using a JEOL JEM-ARM200cF, which is a probe-aberration-corrected microscope with a cold field emission electron gun operating at 200 kV. It is also equipped with a Gatan Quantum electron energy loss spectroscopy (EELS) spectrometer and an Oxford X-max silicon drift detector for energy dispersive spectroscopy (EDS). The cross-sectional TEM specimen of an oxidized REBCO tape was made by mechanical grinding/polishing and Ar ion milling. The cross-sectional TEM samples of oxidized stainless steel was cut by the $Ga^+$ focused ion beam in a Helios NanoLab dual beam scanning electron microscope (SEM). The XPS was performed in a Perkin-Elmer PHI-5100 system with Al $K\alpha$ x-ray energy of 1486.6 eV. The system has an Ar ion milling capability suitable for depth profile analysis. Ar ion source was 3 kV for experiments in this paper. The contact resistivity measurements were performed at both 77 K and 4.2 K under cyclic contact pressure in a previously described facility [7]. For stainless steel tape $\rho_c$ measurements, the sample was sandwiched between two as-received REBCO tapes. So the reported $\rho_c$ values are the sum of two stainless steel/REBCO interfaces.

3. Results and discussions

*3.1 Ebonol® C treatment of REBCO surface*

We measured oxide thickness as a function of treatment time in 20% Ebonol® C solution at 98 °C. The results are plotted in Fig. 1(a). The oxide growth follows a power law of time with an index of 0.6 - 0.8 (the inset of Fig. 1(a)). The original Ebonol® C solution is colorless. It turned to blue color after reaction with copper. This suggests that in the oxidizing process some copper is dissolved in the solution in the form of Cu(II) ions which is blue. This is consistent with the loss of copper also shown in Fig. 1(a). Obviously

Ebonol® C treatment consumes a small fraction of the copper stabilizer. For a 20 µm stabilizer layer with a 1 µm oxide layer, the reduction of the stabilizer is not significantly. The effect of Ebonol® C treatment temperature and solution concentration are shown in Fig. 1(b) and 1(c) respectively. As expected, oxide thickness monotonically increases with both treatment temperature and Ebonol® C concentration.

Fig. 2(a) is a cross-sectional TEM bright field image of the oxide layer on the REBCO tape treated by 20% Ebonol® C solution at 98 °C for 60 s. It shows a rough surface which is slightly less rough than the original plated copper surface of the untreated REBCO tape. There is a fine grain layer of 0.5 - 1 µm on top of the larger grained Cu. Combined elemental EDS maps of oxygen and Cu is shown in Fig. 2(b) confirmed that the fine grain layer is the oxide layer and the larger grains belong to the copper stabilizer. Selected area electron diffraction pattern (Fig. 2(c)) from one grain in the oxide layer was indexed to be a $Cu_2O$ [101]. The extra diffraction spots around the main spots in Fig. 2(c) are from Cu [101] and their double diffractions. This results were reproduced on several other fine grains, which indicates that the oxide layer is mostly $Cu_2O$. The fine structure of Cu $L_{2,3}$ core-loss EELS spectrum from the oxide layer shown in Fig. 2(d) confirms that Cu in the oxide layer is $Cu^{1+}$ [29], which is consistent with the $Cu_2O$ phase.

The Ebonol® C treated sample was also studied by XPS where Cu $2p$ and O $1s$ peaks were scanned after different durations of Ar milling. The Ar milling rate of $Cu_2O$ was not calibrated. After 400 min of Ar milling, sufficiently thick layer of oxide still remained as indicated by strong O $1s$ peak and the fact the surface was still black color. Fig. 3 shows the Cu $2p$ spectra of the oxide layer. The Cu $2p_{3/2}$ peak near 932.5 eV is consistent with that of $Cu_2O$, different from Cu $2p_{3/2}$ of CuO at 933.7 eV [30]. Furthermore, Cu $2p$ of CuO has strong satellite peaks between 943 eV and 962 eV, which is absent in this spectrum. So these results confirm that the oxide layer is $Cu_2O$.

The $\rho_c$ of Ebonol® C treated REBCO tape was measured at 77 K under 25 MPa pressure cycles. The results were presented in our previoius paper [7] which suggest that $\rho_c$ can be controlled by changing treatment time. The $\rho_c$ vs. pressure cycles of a sample treated by Ebonol® C at 98 °C for 1 min was also measured at

4.2 K. The results are shown in Fig. 4. There is a monotonic gradual decrease in $\rho_c$ with number of cycles. At the end of 30,000 cycles, $\rho_c$ is reduced by a factor of 3. The $\rho_c$ of this sample is not temperature sensitive. Because when the temperature increased from 4.2 to 77 K, $\rho_c$ only increased about 5% (data are not shown).

In order to demonstrate the feasibility of large scale oxidation of REBCO tapes, a reel-to-reel Ebonol® C oxidation system was designed and built. A photograph of the system is shown in Fig. 5. The center of the system is an Ebonol® C solution container where oxidization reaction takes place. In addition to the solution temperature control and the tape speed control sub-systems, a water tank and two air nozzle are used to rinse off residual chemicals and dry the tape respectively after the oxidation. Our preliminary results showed that surface preparation, accurate and stable speed and temperature control are the keys in getting uniform copper oxide layer along the length. At the time of writing this paper, total of 400 meter of SuperPower REBCO tapes have been oxidized to be used in various test coils.

### 3.2 Surface oxides on stainless steel co-wind tape

The native oxides on stainless steel surface strongly influence its $\rho_c$. For our samples in as received condition, a $\rho_c$ in the order of 100,000 µΩ-cm² was measured at 77 K under 25 MPa pressure [7]. When they were slightly polished with a Scotch-Brite pad, however, $\rho_c$ reduced dramatically to about 2,400 µΩ-cm². Similarly, when etched by concentrated (37%) HCl for 5 min at room temperature, $\rho_c$ was reduced from dramatically from ~100, 000 to ~1,000 µΩ-cm². Subsequently, the HCl etched samples were stored in air-conditioned lab environment for up to 4 months and their $\rho_c$ were measured at 77 K. The $\rho_c$ versus storage time is plotted in semi-log scale in Fig. 6. Despite the scatter, the data show that $\rho_c$ increased very slowly with storage time. After 4 months of storage, $\rho_c$ was still lower than 3,000 µΩ-cm², much lower than 100,000 µΩ-cm² of the samples in as-received condition. This experiment implies that the native oxides formed in lab environment was very thin and nearly self-passivating. In contrast, the oxide layer formed in the manufacturing environment seems to be significantly thicker.

Apparently in order to increase $\rho_c$ of stainless steel, a thicker oxide layer is needed. We heated a number of samples in air and studied their oxide layer. When heated between 200 and 600 °C, the color of the stainless steel surface changed due to the light interference from the oxides of different thicknesses consistent with reference [31]. $\rho_c$ of the heat treated samples were measured at 77 K under 25 MPa pressure. As shown in Fig. 7, $\rho_c$ increases with heat treatment temperature initially, but seems to level off above 400 °C.

The cross-sections of oxide layers are shown in Fig. 8. It includes annular dark field scanning transmission electron microscopy (ADF-STEM) images and EDS oxygen maps of samples heated for 8 min at 300 °C ((a), (c)) and 1 min at 600 °C ((b), (d)). The oxide layer as indicated by white arrow is about 10 nm and 30 nm thick respectively. The XPS depth profiles for an unheated and a 300 °C 8 min heated samples are shown in Fig. 9 where intensities of Cr $2p$, Fe $2p$ and O $1s$ peaks are plotted against depth. The intensity of each peak is normalized to its maximum value in the depth profile. The depth is obtained by Ar milling rate which is calibrated by TEM. Fig. 9(a) indicates that the native oxide of the unheated sample is only about 3 nm thick, and it consists of mostly Cr oxide. The 300 °C 8 min heated sample (Fig. 9(b)) has about 10 nm thick oxides which contains significant amount of Fe oxides in the top 4 nm then dominated by Cr oxides for the rest of the film.

$\rho_c$ of various samples were measured at 77 K and 4.2 K. Unexpectedly the behavior of $\rho_c$ vs. pressure cycles a 300 °C 8 min heated sample is very different at 4.2 K from that at 77 K as shown in Fig. 10(a). While $\rho_c$ decreases slowly with pressure cycling at 77 K, a dramatic $\rho_c$ reduction of almost 4 orders of magnitude is observed at 4.2 K. This measurement was reproduced 3 times. It is conceivable that for an NI magnet with such a co-wind tape, the magnet properties would change dramatically with load cycles and deviate far from designed values. This would be very problematic. At the end of 30,000 cycles, $\rho_c$ is lower than 100 $\mu\Omega$-cm$^2$. Such a low value confirms that the resistivity and thickness of the oxide layer rather than the bulk resistivity of stainless steel dominates $\rho_c$. Since the contact pressure in an operating NI magnet depends on winding stress, thermal stress, and electromagnetic stress and varies from inner turns to outer turns, it is prudent to investigate the effect of the cycling pressure. Fig. 10(b) compares curves of $\rho_c$ vs. number of cycles of different pressure of 6, 10, and 25 MPa. The $\rho_c$ was measured at the respective maximum pressure. Although $\rho_c$ reduction is less dramatic with lower pressures, it is still over 2 orders of magnitudes after 30,000 cycles.

$\rho_c$ reduction is much less pronounced in the sample heated at 600 °C for 1 min, as shown in Fig. 10(c) which also show that the as-received sample behaved similarly to the 300 °C 8 min heated sample.

*3.3 Discussions*

Due to the considerable uncertainties in the average radius and areal density of asperity spots, it is very difficult to precisely control $\rho_c$. Nevertheless a broad $\rho_c$ control should be possible as demonstrated here by our oxidation experiments of both REBCO surface and the surface of co-winding stainless steel tape. In order to improve the accuracy of $\rho_c$ control, sophisticated surface characterization and thin film deposition techniques seem to be necessary.

The gradual decrease in $\rho_c$ with load cycling due to wear of the oxide layer is expected as we observed in Ebonol® C treated REBCO tapes. The dramatic reduction of $\rho_c$ at 4.2 K with pressure cycles in stainless steel tape heated at 300 °C can be also attributed to the wear of the oxide layer. After pressure cycling, the surface of this sample was examined by SEM. Neither the contact spots nor any evidence of surface wear were discernable. A 20 μm long cross-sectional TEM specimen was also made from this sample. The evidence of surface wear was not found either. The difference in cycling behavior of this sample between 77 K and 4.2 K, however, is surprising. It may be speculated that higher thermal stress caused by differential thermal contraction between the oxide and stainless steel at 4.2 K makes the asperity spots more vulnerable to wear. In practice, this means that pressure cycling tests at 77 K does not correlate well with the 4.2 K behavior. So for developing a resistive coating for NI magnets operating at 4.2 K, $\rho_c$ under cyclic pressure might have to be tested at 4.2 K. This dramatic $\rho_c$ reduction did not occur to the sample with about 30 nm oxide formed at 600 °C. This suggests that thicker oxide layer has better wear resistance. It seems that a wear-resistant oxide layer needs to be thicker than 10 nm. Since a 10 nm oxides corresponds to ~ 100,000 µΩ-cm$^2$ in $\rho_c$, a thicker oxide layer would result in a $\rho_c$ of greater than 100,000 µΩ-cm$^2$. Therefore $\rho_c$ lower than that will be difficult to obtain without suffering from 4.2 K load cycling sensitivity.

## 4  Conclusions

The control of $\rho_c$ is critical to NI REBCO coil technology. One way to control $\rho_c$ is by controlling the thickness and resistivity of the surface oxides of conductor or co-wind tapes. We used a commercial oxidizing agent Ebonol® C to treat the copper surface of REBCO tapes. The copper oxide layer was characterized by TEM and XPS. The oxide layer grown in Ebonol® C at 98 °C for 1 min is about 0.5 µm of Cu$_2$O. The $\rho_c$ between two oxidized REBCO is in the order of 35 mΩ-cm$^2$ at 4.2 K which decreases slowly with contact pressure cycles. The $\rho_c$ increases but only slightly at 77 K. We also investigated the effect of oxidation of stainless steel co-wind tape on $\rho_c$ in order to control it. The native oxides on 316 stainless steel tape as well as those heated in air at 200 - 600 °C were examined by TEM and XPS. The native oxides layer is mostly Cr oxides of about 3 nm thick. After heating at 300 °C for 8 min and 600 °C for 1 min, the oxides becomes mostly Fe oxide and its thickness increases to about 10 and 30 nm respectively. For the stainless steel tapes with about 10 nm surface oxides, pressure cycling for 30,000 cycles decreases $\rho_c$ by almost 4 orders of magnitude. Whereas at 77 K, it only changes slightly. For a surface with 30 nm oxide, the $\rho_c$ decreases moderately with load cycles. The results suggest that for an oxidized stainless steel to achieve stable $\rho_c$ over large number of load cycles a relatively thick oxide film is needed.

## 5  Acknowledgement

We thank Robert E Goddard, Robert P Walsh and Dr. Ke Han for assistance in our experiments. The NHMFL is supported by NSF through NSF-DMR-1157490 and 1644779, and the State of Florida.

Captions

Fig. 1. The thickness of $Cu_2O$ on Cu stamp samples treated in Ebonol® C special solution (a) as a function of treatment time at 98 °C in 20% solution. The copper thickness loss is also plotted. The inset is a log-log plot of the same data. (b) as a function of reaction temperature for 1 min. treatment in 20% solution (c) as a function of concentration at 98 °C for 1 min treatment.

Fig. 2 Cross-sectional TEM results of the oxide layer on a REBCO tape treated by Ebonol® C at 98 °C for 1 min. (a) bright field image, the thin dashed line indicate the boundary between the fine grained top layer and the large gained substrate. (b) A combined EDS elemental map of Cu (orange) and O (blue) indicating that the top fine grained layer is copper oxide. (c) The electron diffraction pattern from a grain in the oxide layer. The main spots are from $Cu_2O$ [101]. The extra spots are from Cu [101] and the double diffractions. (d) EELS spectrum of Cu $L_{2,3}$ core-loss from the oxide layer, which is consistent with $Cu_2O$ spectrum.

Fig. 3. XPS Cu $2p$ spectra of REBCO tape surface oxidized by Ebonol® C for 1 min at 98 °C. The surface Ar milling time is indicated at each spectrum. The vertical dashed lines indicate the peak positions of Cu $2p$ of $Cu_2O$ and CuO according to ref. [30].

Fig. 4. $\rho_c$ versus pressure cycles of a REBCO tape treated by Ebonol® C for 1 min at 98 °C measured at 4.2 K under 25 MPa pressure. The cyclic contact pressure was 2.5 – 25 MPa.

Fig. 5. The reel-to-reel Ebonol® C oxidation treatment system.

Fig. 6. $\rho_c$ of a 316 stainless steel tape as a function of exposure time in laboratory environment after 5 min etch by 37% HCl. In the measurement, the stainless steel sample was sandwiched between two REBCO tapes. $\rho_c$ was measured at 77 K under 25 MPa contact pressure after 5 pressure cycles of 2.5 – 25 MPa.

Fig. 7. $\rho_c$ of stainless steel tape samples at 77 K under 25 MPa pressure as a function of heat treatment temperature. The heat treatment time was 1 min.

Fig. 8 Cross-sectional ADF-SETM images of oxide layers on 316 stainless steel. TEM sample was made by focused ion beams. A Pt capping layer was deposited on surface to protect the oxide layer during the sample making process (a) heat treated at 300 °C for 8 min. The white arrow points to the oxide layer (b) heat treated at 600 °C for 1 min. (c) an EDS oxygen map of sample (a). (d) an EDS oxygen map of the sample (b).

Fig. 9. XPS depth profile of surface oxides of 316 stainless steel. Cr 2$p$, Fe 2$p$ and O 1$s$ were analyzed (a) native oxide, after a few days after 37% HCl etching for 5 min, (b) oxidized by heating at 300 °C for 8 min. The Ar milling rate was calibrated against the thickness measured by TEM.

Fig. 10. The effect of contact load cycling on $\rho_c$ of 316 stainless steel. (a) the sample was heated at 300 °C for 8 min measured at 77 and 4.2 K. (b) The effect of maximum cycling pressures. $\rho_c$ was measured at the corresponding maximum pressure at 4.2 K. (c) The $\rho_c$ cycling effect of different surface conditions, measured at 4.2 K under 25 MPa pressure.

Fig. 1

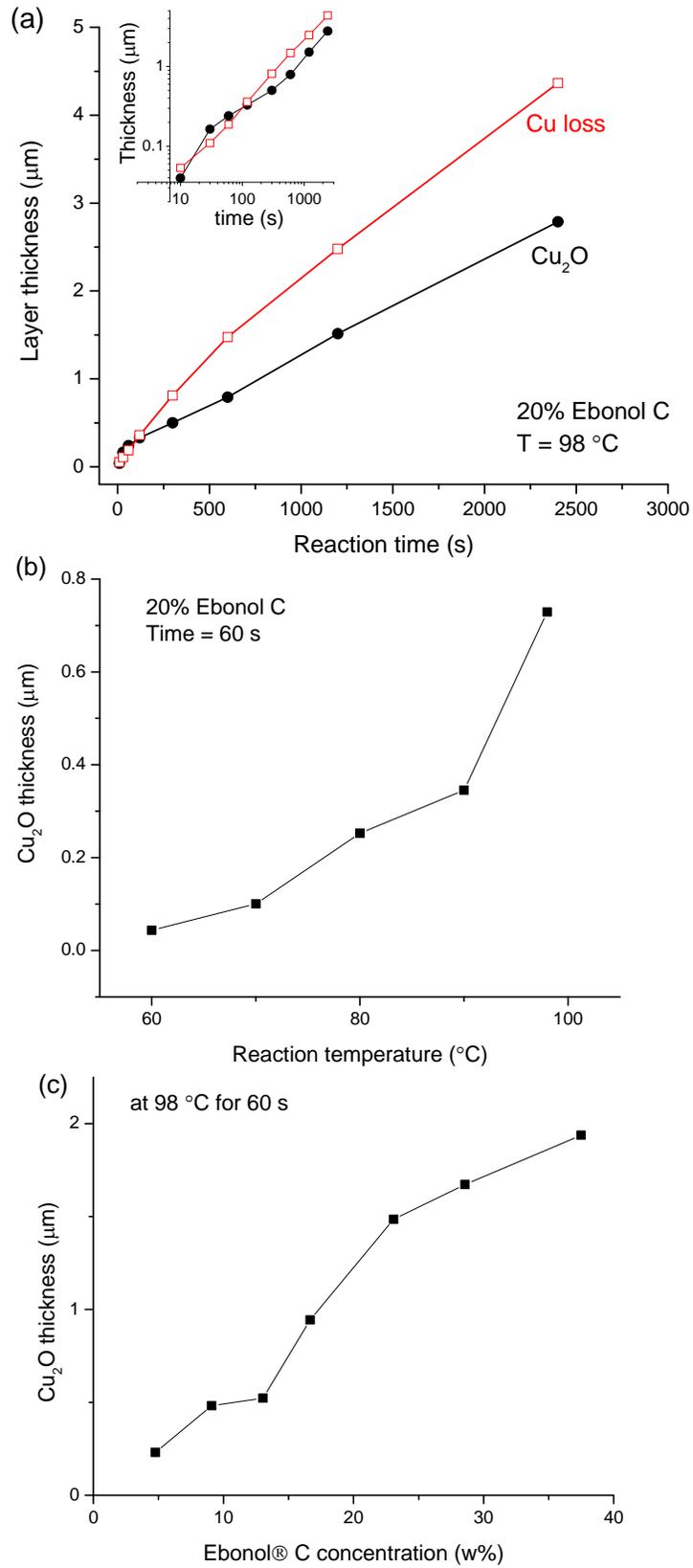

Fig. 2

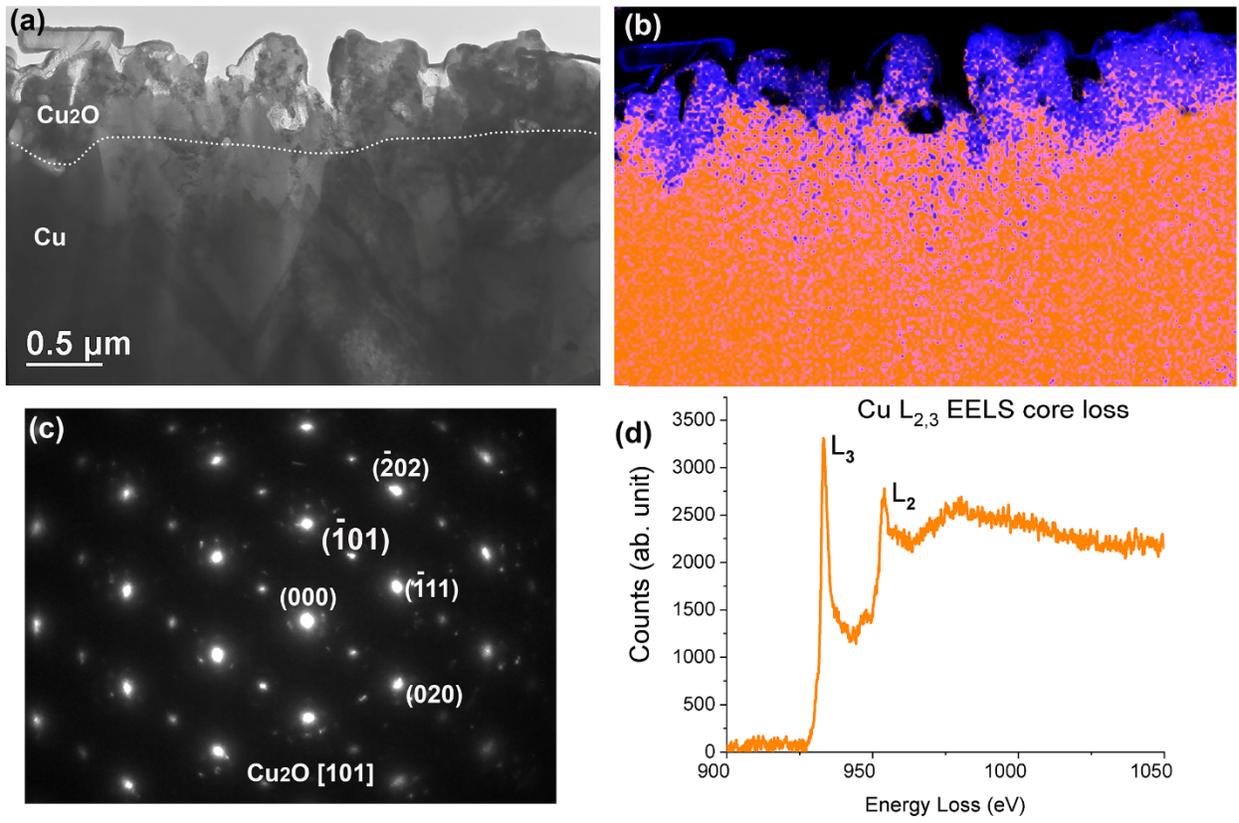

Fig. 3

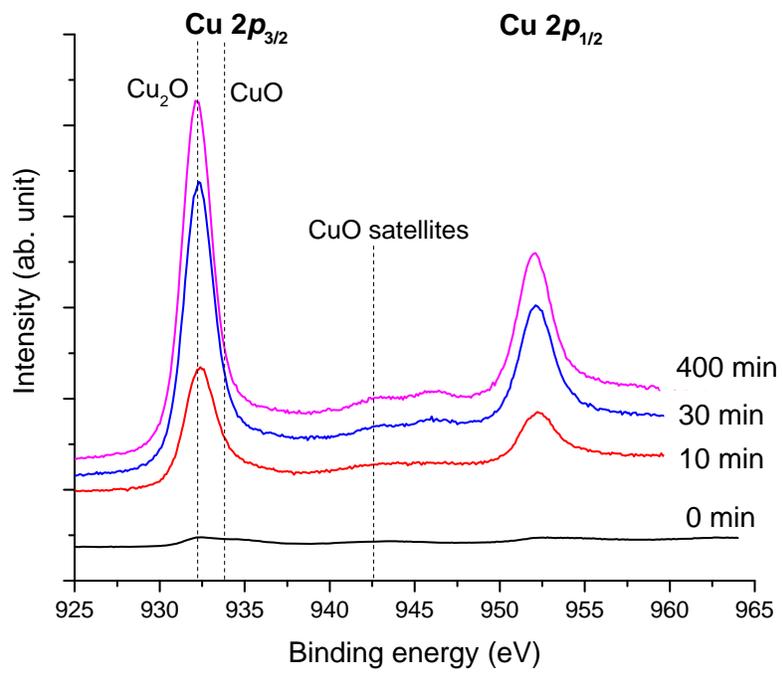

Fig. 4

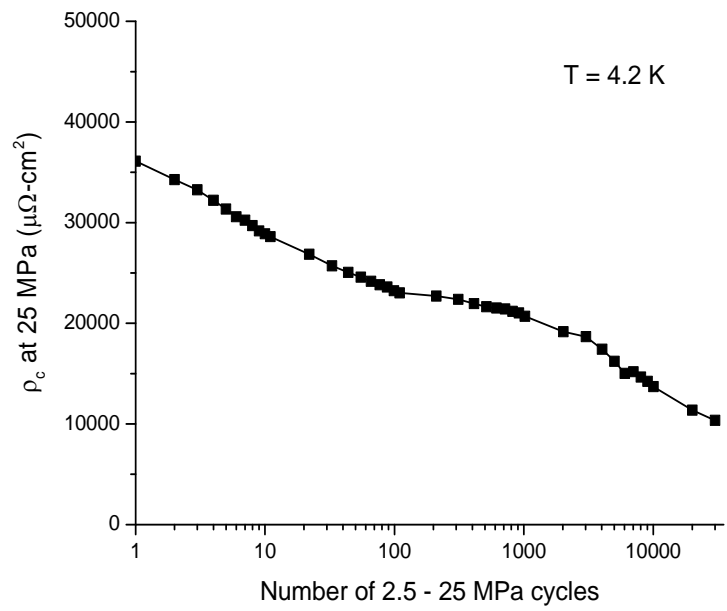

Fig. 5

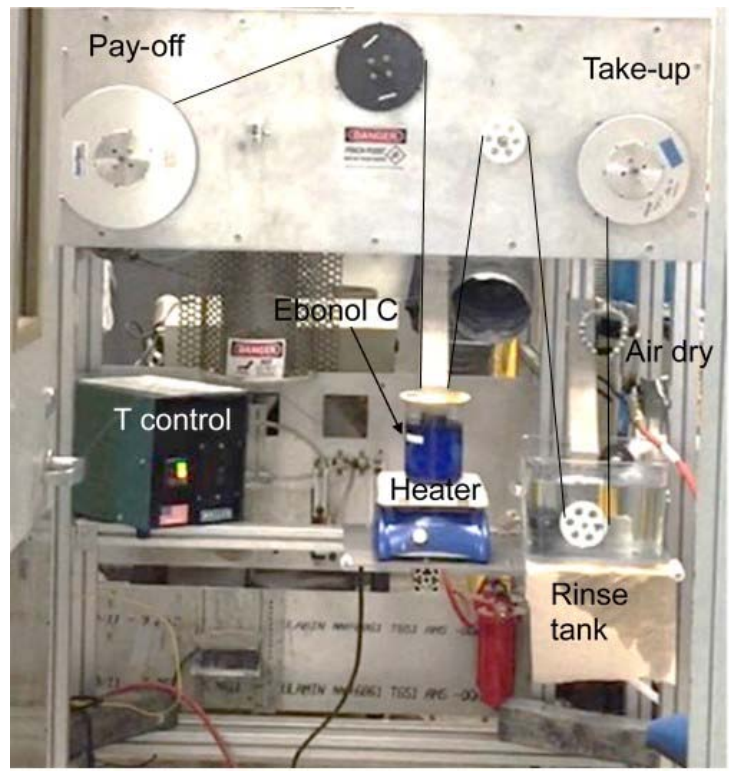

Fig. 6

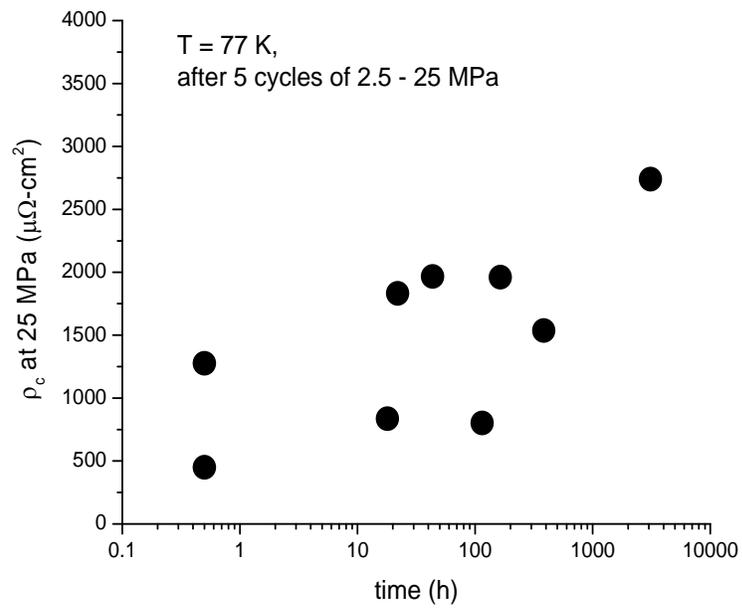

Fig. 7

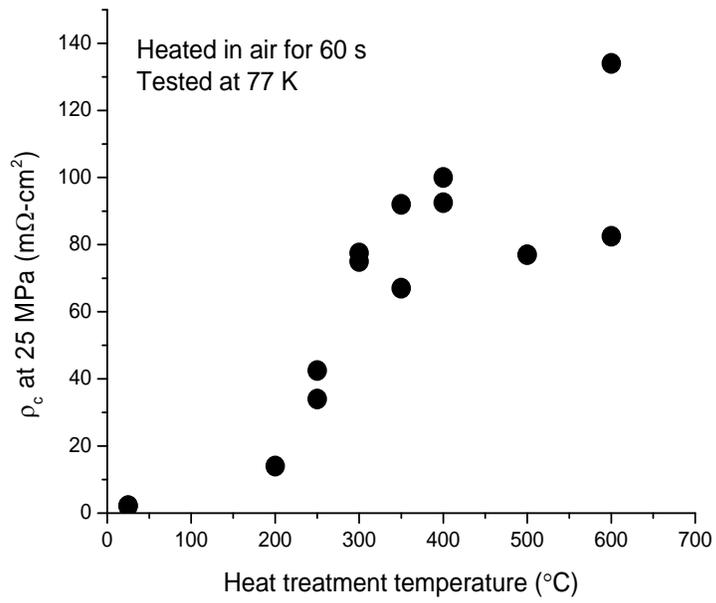

Fig. 8

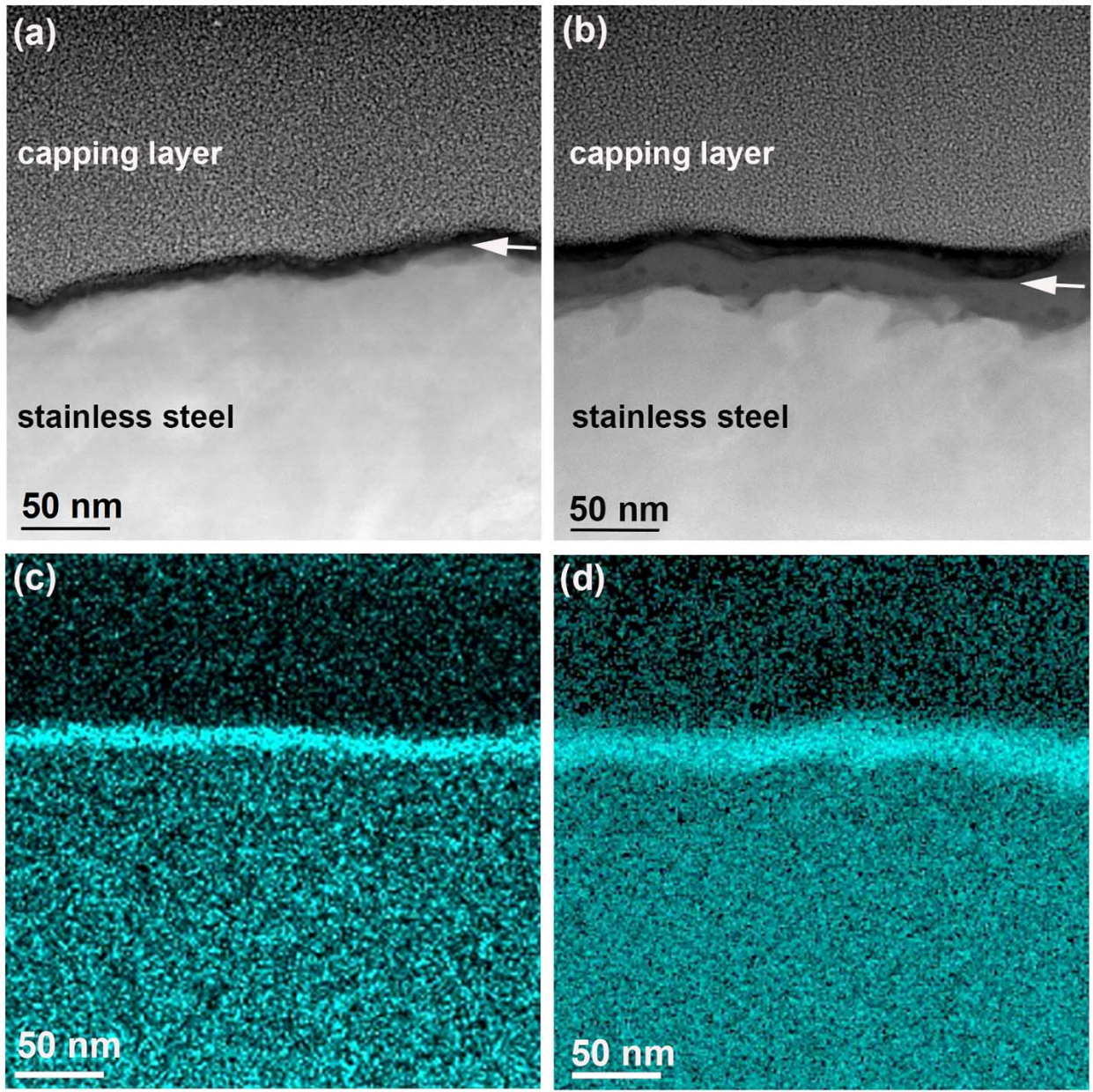

Fig. 9

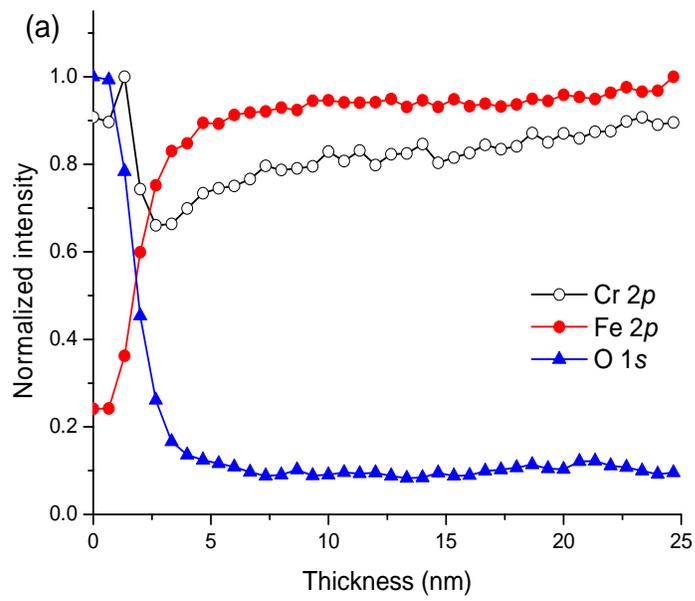

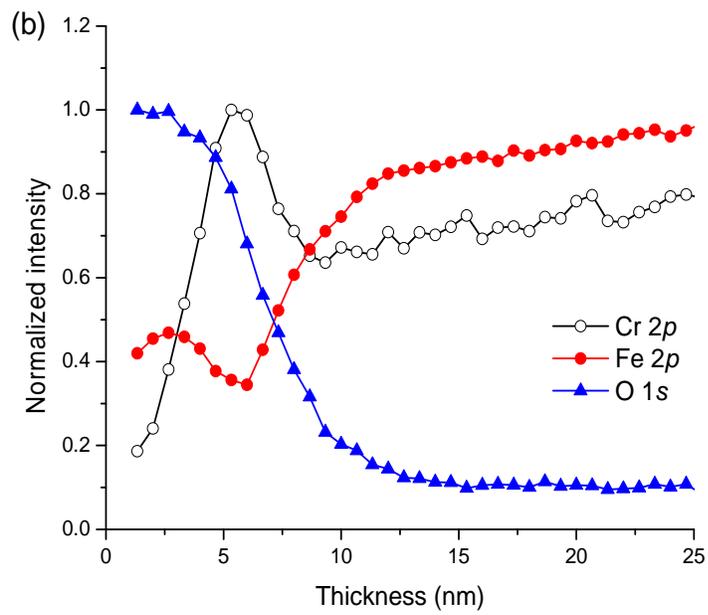

Fig. 10

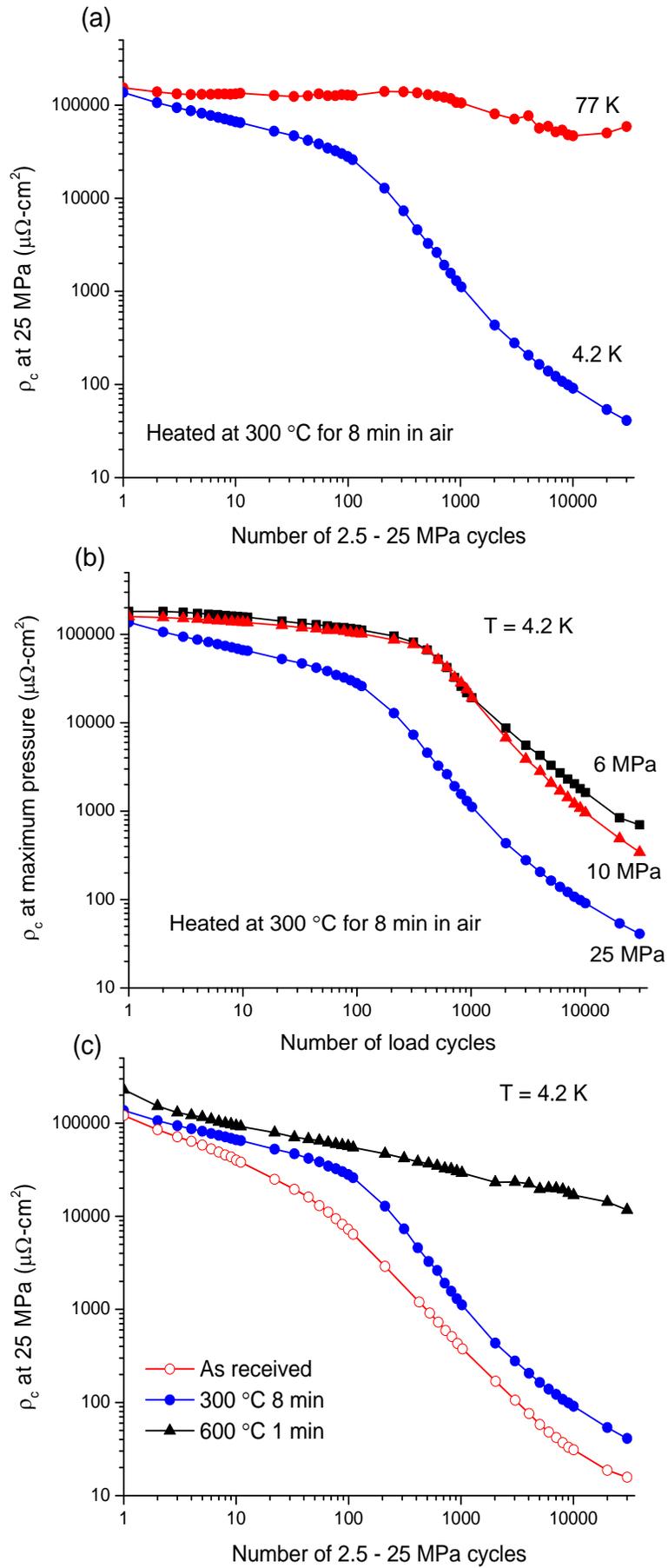